\documentclass[12pt]{article}

\usepackage{amssymb}
\usepackage{amsmath}
\usepackage{amscd}
\usepackage{latexsym}
\usepackage{graphicx}

\usepackage{cite}

\newcommand{\be}{\begin{equation}}
\newcommand{\ee}{\end{equation}}

\newcommand{\dlt}{\delta}

\newcommand{\bt}{\beta}
\newcommand{\vp}{\varphi}
\newcommand{\ep}{\varepsilon}
\newcommand{\al}{\alpha}
\newcommand{\ra}{\rightarrow}

\newcommand{\gm}{\gamma}
\newcommand{\om}{\omega}

\begin{document}

\begin{center}

{\Large{\bf Self-similar sequence transformation for critical exponents} \\ [5mm]

V.I. Yukalov$^{1,2}$ and E.P. Yukalova$^{3}$ }  \\ [3mm]

{\it
$^1$Bogolubov Laboratory of Theoretical Physics, \\
Joint Institute for Nuclear Research, Dubna 141980, Russia \\ [2mm]

$^2$Instituto de Fisica de S\~ao Carlos, Universidade de S\~ao Paulo, \\
CP 369, S\~ao Carlos 13560-970, S\~ao Paulo, Brazil \\ [2mm]

$^3$Laboratory of Information Technologies, \\
Joint Institute for Nuclear Research, Dubna 141980, Russia } \\ [3mm]

{\bf E-mails}: {\it yukalov@theor.jinr.ru}, ~~ {\it yukalova@theor.jinr.ru}

\end{center}

\vskip 1cm

\begin{abstract}

Self-similar sequence transformation is an original type of nonlinear sequence 
transformations allowing for defining effective limits of asymptotic sequences.
The method of self-similar factor transformations is shown to be regular. This 
method is applied for calculating the critical exponents of the $O(N)$-symmetric 
$\varphi^4$ theory in three dimensions by summing asymptotic $\ep$ expansions. 
It is shown that this method is straightforward and essentially simpler than 
other summation techniques involving complicated numerical calculations, while 
enjoying comparable accuracy.    
 
\end{abstract}

\vskip 2mm

{\bf Keywords}: Self-similar sequence transformation; asymptotic series; 
summation methods; critical exponents

\vskip 3mm

\section{Introduction}

Asymptotic expansions in powers of some parameter are widely used in physics and
applied mathematics \cite{Giacaglia_1,Nayfeh_2}. Since the parameters of interest 
very rarely are really small, one needs to employ some kind of effective summation 
of divergent series. The most popular are the method of Pad\'{e} approximants 
\cite{Baker_3}, Borel summation \cite{Kleinert_4} and its variants, such as 
Pad\'{e}-Borel summation and Borel summation with conformal mapping. One also uses 
the methods of renormalization group, conformal bootstrap, Monte Carlo simulations, 
and other methods (see review \cite{Dupuis_5}) requiring quite heavy numerical 
calculations. 

In the present paper, we advocate another approach based on self-similar approximation 
theory \cite{Yukalov_6,Yukalov_7,Yukalov_8,Yukalov_9}. The idea of this theory is to
represent the transition from one approximation term to another as the motion of a 
dynamical system, or a renormalization-group equation, with the approximation order 
playing the role of discrete time. Then the sequence of approximation terms becomes 
bijective to the dynamical-system trajectory and the limit of the sequence is bijective 
to the fixed point of the trajectory. In the vicinity of a fixed point, the evolution 
equation (renormalization-group equation) acquires the form of a self-similar relation,
which explains the name of the self-similar approximation theory. Mathematical details
can be found in the review articles \cite{Yukalov_10,Yukalov_11}. 

The approach has been applied to numerous problems, providing good agreement with the 
exact results, when these are available, as well as with numerical calculations and 
other elaborate methods, thus being compatible with other methods of summation of
divergent series. Accurate results can be obtained even when just a few terms of an 
expansion are given, and when other summation methods are not applicable at all 
(see reviews \cite{Yukalov_10,Yukalov_11}). 

The method demonstrates good numerical convergence, which becomes especially evident 
for the cases, where a large number (of order or larger than ten) of perturbative 
terms are available. This concerns, e.g., the so-called zero-dimensional model 
\cite{Yukalov_12}, one-dimensional anharmonic oscillator \cite{Yukalov_12}, and spin 
glass \cite{Yukalov_13}.  

All one needs for the application of the method is an asymptotic expansion centered 
at a point on the real axis. Its analytical behavior on the whole complex plane is 
not required. Thus, in the case of the zero-dimensional model and the one-dimensional 
oscillator, we meet expressions that have on the complex plane a singular point at 
zero \cite{Bender_14}. This does not hinder the use of the method resulting in 
numerically convergent sequence of approximants for these models \cite{Yukalov_12}.

The method allows for the summation of a large class of functions, rational, irrational 
and transcendental. Moreover, there exists a class of functions that are exactly 
reproducible by this method \cite{Yukalov_12}. This is the class of functions having 
the form
$$
 F_{k_M}(x) = \prod_{i=1}^M P_{m_i}^{\al_i}(x)  
$$
of the product of polynomials 
$$
P_{m_i}(x) = c_{i0} + c_{i1} x + c_{i2} x^2 + \ldots + c_{i_{m_i}}x^{m_i} \,
$$
where $\al_i$ and  $c_{ij}$ are either real or the powers $\al_i$ and coefficients 
$c_{ij}$ are complex-valued numbers entering in complex conjugate pairs so that $F_{k_M}$ 
is real, and
$$
k_M = \sum_{i=1}^M m_i + M \;  .
$$
The exponential function also is shown to be reconstructed exactly starting from 
the second-order approximation \cite{Yukalov_12,Yukalov_15}.

Nonlinear differential equations, including singular equations, can be solved by this
method, first, by deriving a solution in terms of a series in powers of a variable, 
and then by summing this series using the self-similar factor transformation. For 
some nonlinear equations, exact soliton solutions have been obtained 
\cite{Yukalov_12,Yukalova_16}. 

In the present paper, we consider two important points, one technical and the other
of great interest in physics. The first point is the proof of the method regularity.  
We consider the method of self-similar factor transformation and show that this method 
is regular, which implies that it sums every convergent series to the same sum as that 
to which the series converges. This point is of high importance for the justification 
of the approach. It is the standard way of dealing with summation methods, when one, 
first, shows the regularity of the method and then extrapolates it to divergent 
sequences, by demonstrating its compatibility with other reliable methods and observing
numerical convergence for some test problems \cite{Hardy_17,Weniger_18}. Then we apply
the method to the transformation of $\varepsilon$ expansions for critical exponents. 
As a concrete example, we study the $O(N)$ -symmetric $\varphi^4$ theory in three 
dimensions. We show that the method of self-similar factor transformation provides 
the accuracy comparable with other elaborate techniques involving heavy numerical 
calculations, while being essentially simpler.  
 
In Sec. 2, we formulate the method of self-similar factor transformation. We do not 
plunge into the foundations of the whole theory, but will just give the receipt of 
the method usage. In Sec. 3, we prove the regularity of the method. The application 
to defining the critical exponents is given in Sec. 4, where we compare our results 
with the most accurate values obtained by other methods summarized in Refs.
\cite{Dupuis_5,Pelissetto_19,Abhignan_20,Shalaby_21}. We compare our results with 
Monte Carlo simulations, conformal bootstrap, hypergeometric Meijer summation, Borel 
summation, Borel summation with conformal mapping, and the method of nonperturbative 
renormalization group. This comparison shows good agreement of the self-similar factor 
approximants with the calculations by other methods. The last Sec. 5 concludes.

\section{Self-similar factor transformation}

In this section, we describe the method that can be used for the summation of arbitrary 
asymptotic series. Suppose we have got an asymptotic expansion for a real function
\be
\label{1}
f_k(x) = f_0(x) \left( 1 + \sum_{n=1}^k a_n x^n \right)
\ee
in powers of a real parameter $x$ assumed to be asymptotically small. However, we need 
to find the value of the function at a finite value of the parameter. The extrapolation 
of the expansion $f_k(x)$ to arbitrary values of the parameter $x$ can be done by means of 
the self-similar factor transformation \cite{Yukalov_10,Yukalov_22,Gluzman_23}. This 
method transforms the truncated expansion (\ref{1}) to the factor form 
\be
\label{2}
 f_k^*(x) = f_0(x) \prod_{j=1}^{N_k} ( 1 + A_j x )^{n_j} \;  ,
\ee
where the number of factors is
\begin{eqnarray}
\label{3}
N_k = \left\{ \begin{array}{rl}
k/2 ,     ~ & ~ k = 2, 4, 6, \ldots \\
(k+1)/2 , ~ & ~ k = 3, 5, 7, \ldots \end{array}
\right.   .
\end{eqnarray}
The parameters $A_j$ and $n_j$ are uniquely determined by the accuracy-through-order 
procedure, by equating the like-order terms in the expansions at small $x$,
\be
\label{4}
f_k^*(x) \simeq f_k(x) \qquad ( x \ra 0 ) \;   .
\ee
This procedure gives the equations
\be
\label{5}
 \sum_{j=1}^{N_k} n_j A_j^n = D_n \qquad ( n = 1, 2, \ldots , k)  \; ,
\ee
in which
\be
\label{6}
 D_n\equiv \frac{(-1)^{n-1}}{(n-1)!} \; \lim_{x\ra 0} \;
\frac{d^n}{dx^n}\; \ln \left( 1 + \sum_{m=1}^n a_m x^m \right) \;  .
\ee

In the case of an even order $k$, Eq. (\ref{5}) consists of $k$ equations uniquely 
defining the $k/2$ parameters $A_j$ and $k/2$ parameters $n_j$. For odd orders $k$, 
the system of $k$ equations (\ref{5}) contains $k+1$ unknowns, where one of the 
parameters $A_j$, say $A_1$, is arbitrary. Normalizing $A_j$ in units of $A_1$ 
implies $A_1=1$, which makes the system of equations (\ref{5}) self-consistent and 
all parameters uniquely defined \cite{Yukalov_10,Yukalov_15}. If the found parameters 
lead to a complex-valued approximant, it is replaced by the nearest real-valued 
approximant. The final result is given by the average between the last two approximants 
$[f_k^*(x)+f_{k-1}^*(x)]/2$ and the error bar is defined as the half-difference between 
the last two different approximants $[f_k^*(x)-f_{k-1}^*(x)]/2$. As is seen, the scheme 
is very simple and straightforward.

\section{Method regularity} 

In this section, we show that the self-similar factor transformation is a regular method
of summation. 

\vskip 2mm

{\bf Theorem}. Let us consider the sequence of the terms
\be
\label{7}
f_k(x) = \sum_{n=0}^k a_n x^n  \qquad ( x \in \mathbb{D} )
\ee
that are defined on a domain $\mathbb{D}\subset\mathbb{R}$ including the point $x=0$. 
The self-similar factor transformation reduces the term $f_k(x)$ to the form 
\be
\label{8}
f_k^*(x) \equiv a_0 \prod_{j=1}^{N_k} ( 1 + A_j x)^{n_j}
\ee
that is a smooth function on $\mathbb{D}$, hence being infinitely differentiable, whose 
parameters $A_j$ and $n_j$ are prescribed by the accuracy-through-order procedure, such 
that
\be
\label{9}
\frac{f_k^{*(n)}(0)}{n!} = a_n \qquad ( n = 0,1,2,\ldots) \;  , 
\ee
where $f_k^{*(n)}(x)$ is the $n$-th derivative of $f_k^*(x)$ over $x$.

If the sequence of terms (\ref{7}) converges to a smooth function
\be
\label{10}
 f(x) = \sum_{n=0}^\infty a_n x^n  \; ,
\ee
then the sequence of the factor approximants (\ref{8}) converges to the function 
\be
\label{11}
f^*(x) \equiv a_0 \prod_{j=1}^\infty ( 1 + A_j x)^{n_j}
\ee
coinciding with $f(x)$: 
$$
f^*(x) = f(x) \; .
$$

\vskip 2mm

{\it Proof}. The convergence of the sequence of terms (\ref{7}) to a smooth function
$f(x)$ implies that the latter can be represented as the Taylor series
\be
\label{12}
 f(x) = \sum_{n=0}^\infty \frac{f^{(n)}(0)}{n!}\; x^n \;  ,
\ee
with
\be
\label{13}
\frac{f^{(n)}(0)}{n!} = a_n \qquad ( n = 0,1,2,\ldots )    
\ee
and with the remainder
\be
\label{14}
R_k[\; f(x) \; ] = \frac{f^{(k+1)}(c_1)}{(k+1)!} \; x^{k+1} \;  ,
\ee
where $c_1 \in (0,x)$, this remainder tending to zero when $k \ra \infty$,
\be
\label{15}
\lim_{k\ra\infty} \; R_k[\; f(x) \; ] = 0 \;  .
\ee

Comparing expressions (\ref{9}) and (\ref{13}), we see that
\be
\label{16}
f^{(n)}(0) - f_n^{*(n)}(0) = 0 \qquad ( n = 0,1,2,\ldots ) \;  .
\ee
Because of this, and since the derivatives are continuous, there exists a finite 
value $\varepsilon$ such that
\be
\label{17}
| \; f^{(n)}(c_1) - f_n^{*(n)}(c_2)\; | < \ep < \infty  \; ,
\ee
where $0 \leq c_2 \leq x$ and $n = 0,1,2,\ldots$. Considering the difference between 
the remainder (\ref{14}) and the remainder
\be
\label{18}
 R_k[\; f_k^*(x) \; ] = \frac{f_k^{*(k+1)}(c_2)}{(k+1)!} \; x^{k+1} \;  ,
\ee
we have
\be
\label{19}
 |\; R_k[\; f_k^*(x) \; ] - R_k[\; f(x) \; ] \; | < 
\frac{\ep}{(k+1)!} \; |\; x\; |^{k+1} \; .
\ee
Using the Stirling formula $n! \simeq \sqrt{2 \pi n}(n/e)^n$, we find that
$$ 
\lim_{n\ra\infty}\; \frac{\ep}{(n+1)!} \; |\; x\; |^{n+1}  = 
\frac{\ep}{\sqrt{2\pi}} \; \lim_{n\ra\infty}\; 
\left( \frac{|\; x\;|\; e}{n}\right)^n = 0
$$ 
for any fixed $x$. From here, and taking into account the limit (\ref{15}), we get
\be
\label{20}
 \lim_{k\ra\infty}\; R_k[\; f_k^*(x) \; ] = 0 \; .
\ee
Then, by the Taylor theorem, the factor approximant (\ref{11}) can be represented 
in the form of the Taylor series
\be
\label{21}
 f^*(x) = \sum_{n=0}^\infty \frac{f^{*(n)}(0)}{n!}\; x^n \; , 
\ee
with 
\be
\label{22}
\frac{f^{*(n)}(0)}{n!} = a_n \qquad ( n = 0,1,2,\ldots) \; . 
\ee
Comparing series (\ref{12}) and (\ref{21}), under equalities (\ref{13}) and (\ref{22}), 
we come to the conclusion that the functions $f^*(x)$ and $f(x)$ coincide. $\square$

\section{Critical exponents}

Critical exponents can be presented in the form of $\varepsilon$-expansions in 
powers of $\ep=4-d$, where $d$ is space dimensionality. For $O(N)$-symmetric 
$\vp^4$ theory in three dimensions, the five-loop expansions can be found in the 
book \cite{Kleinert_24}. The summation of the five-loop expansions by means 
of self-similar approximants was considered in Ref. \cite{Yukalov_13} for all $N$. 
It was shown that for $N=-2$ and $N \ra \infty$ self-similar approximations yield 
the exact values for the exponents. The results are very accurate for large $N\gg 1$, 
with the errors decreasing as $1/N$ with increasing $N$. However the accuracy for the 
lower $N$ was not sufficient. 

Our aim in the present paper is to demonstrate that the accuracy of self-similar 
factor approximants for the $O(N)$ -symmetric $\varphi^4$ theory in three dimensions 
can be drastically improved by employing the available seven-loop $\varepsilon$-expansions 
that are known for $N = 1$ \cite{Ryttov_25} and are derived, using the seven-loop 
coupling parameter expansions \cite{Schnetz_26}, in Ref. \cite{Shalaby_27} for 
$N = 0,1,2,3,4$. These expansions are as follows.

\vskip 3mm
(i) $N = 0$. 

\vskip 2mm
For $N = 0$, we have
\be
\label{23}
\nu^{-1} = 2 - 0.25\;\ep - 0.08594 \;\ep^2 + 0.11443 \;\ep^3 - 0.28751 \;\ep^4 + 
 0.95613 \;\ep^5 - 3.8558 \;\ep^6 + 17.784 \;\ep^7  \; ,
\ee
\be
\label{24}
\eta = 0.015625\;\ep^2 + 0.016602\; \ep^3 - 0.0083675\; \ep^4 + 0.026505\; \ep^5 - 
 0.09073\; \ep^6 + 0.37851\; \ep^7 \; ,
\ee
and
\be
\label{25}
\om = \ep - 0.65625\; \ep^2 + 1.8236\; \ep^3 - 6.2854\; \ep^4 + 26.873\; \ep^5 - 
 130.01\; \ep^6 + 692.1\; \ep^7 \; .
\ee
   
We calculate the corresponding self-similar factor approximants $f_k^*(\ep)$, as
is explained in the previous section, and set $\varepsilon = 1$. The results 
of this calculation for the exponent $\nu$ are illustrated in Table 1. Following 
the same procedure for the exponents $\eta$ and $\om$, we obtain the values shown 
in Table 2, where we compare the results obtained by means of factor approximants 
(FA) with those of other methods: Monte Carlo simulations (MC) 
\cite{Hasenbusch_28,Hasenbusch_29,Hasenbusch_30,Clisby_31,Clisby_32,Hasenbusch_33}, 
Conformal bootstrap (CB) \cite{Showk_34,Shimada_35,Kos_36,Echeverri_37,Simmons_38}, 
Hypergeometric Meijer summation (HGM) \cite{Shalaby_27}, Borel summation complimented
by additional conjectures on the behavior of coefficients (BAC) \cite{Kompaniets_39}, 
Borel summation with conformal mapping (BCM) \cite{Kompaniets_40}, and Nonperturbative 
renormalization group (NPRG) \cite{Dupuis_5,Hasselmann_41,Rose_42,Polsi_43}.      

\vskip 3mm
(ii) $N=1$. 

\vskip 2mm
The $\ep$ expansions for the critical exponents read as
\be
\label{26}
\nu^{-1} = 2 - 0.333333\;\ep - 0.11728\; \ep^2 + 0.12453\; \ep^3 - 
0.30685\; \ep^4 +  0.95124\; \ep^5 - 3.5726\; \ep^6 + 15.287\; \ep^7  \; ,
\ee
\be
\label{27}
\eta = 0.018519\; \ep^2 + 0.01869\; \ep^3 - 0.0083288\; \ep^4 + 0.025656\; \ep^5 - 
 0.081273\; \ep^6 + 0.31475\; \ep^7 \; ,
\ee
and
\be
\label{28}
\om = \ep - 0.62963\; \ep^2 + 1.6182\; \ep^3 - 5.2351\; \ep^4 + 20.75\; \ep^5 - 
93.111\; \ep^6 + 458.74\; \ep^7 \; .
\ee

The results for the factor approximants are summarized in Table 3, where they are 
compared with the values obtained by other methods listed above.  

\vskip 3mm
(iii) $N=2$.

\vskip 2mm
The $\varepsilon$ expansions are
\be
\label{29}
\nu^{-1} =  2 - 0.4\; \ep - 0.14\;\ep^2 + 0.12244\; \ep^3 - 0.30473\; \ep^4 + 
0.87924\; \ep^5 -  3.103\; \ep^6 + 12.419\; \ep^7 \; ,
\ee
\be
\label{30}
\eta = 0.02\; \ep^2 + 0.019\; \ep^3 - 0.0078936\; \ep^4 + 0.023209\; \ep^5 - 
0.068627\; \ep^6 +  0.24861\; \ep^7 \; ,
\ee
and
\be
\label{31}
\omega = \ep - 0.6\; \ep^2 + 1.4372\; \ep^3 - 4.4203\; \ep^4 + 16.374\; \ep^5 
- 68.777\; \ep^6 + 316.48\; \ep^7 \; .
\ee 

The calculated factor approximants for the critical exponents are presented in 
Table 4, where they are compared with the exponents found by other methods listed 
above. 

\vskip 3mm
(iv) $N=3$.

\vskip 2mm
The $\varepsilon$ expansions read as
\be
\label{32}
\nu^{-1} =  2 - 0.45455\; \ep - 0.1559\; \ep^2 + 0.11507\; \ep^3 - 
0.2936\; \ep^4 + 0.78994\; \ep^5 - 2.6392\; \ep^6 + 9.9452\; \ep^7 \; ,
\ee
\be
\label{33}
\eta = 0.020661\; \ep^2 + 0.018399\; \ep^3 - 0.0074495\; \ep^4 + 
0.020383\; \ep^5 -  0.057024\; \ep^6 + 0.19422\; \ep^7 \; ,
\ee
and
\be
\label{34}
\om = \ep - 0.57025\; \ep^2 + 1.2829\;\ep^3 - 3.7811\;\ep^4 + 13.182\;\ep^5 - 
 52.204\;\ep^6 + 226.02\;\ep^7  \; .
\ee 

The corresponding exponents calculated by means of the factor approximants are shown in 
Table 5, together with the results of calculation by other methods listed above.

\vskip 3mm
(v) $N=4$. 

\vskip 2mm
The $\varepsilon$ expansions take the form
\be
\label{35}
\nu^{-1} =  2 - 0.5\; \ep - 0.16667\; \ep^2 + 0.10586\; \ep^3 - 0.27866\; \ep^4 + 
0.70217\; \ep^5 -  2.2337\; \ep^6 + 7.9701\; \ep^7 \; ,
\ee
\be
\label{36}
\eta = 0.020833\; \ep^2 + 0.017361\; \ep^3 - 0.0070852\; \ep^4 + 0.017631\; \ep^5 - 
 0.047363\; \ep^6 + 0.15219\; \ep^7 \; ,
\ee
and
\be
\label{37}
\omega = \ep - 0.54167\; \ep^2 + 1.1526\; \ep^3 - 3.2719\; \ep^4 + 
10.802\; \ep^5 - 40.567\;\ep^6 + 166.26\; \ep^7 \; .
\ee 
 
The resulting values of the factor approximants and the values of the exponents 
found by other methods listed above are given in Table 6. 

As has been explained above, the construction of the factor approximants for 
each given expansion is straightforward. In order that the reader would grasp the 
overall structure of these approximants, we adduce below, as an example, the explicit 
expressions for $\nu^{-1}$ in the case of $N=0$. The self-similar approximant of order 
$k=2$ is
$$
f_2^*(\ep) = 2(1 - 0.81252\ep)^{0.153842} \;  .
$$ 
For the approximant of order $k=3$, we have
$$
f_3^*(\ep) =  \frac{2(1 + \ep)^{0.169077}}{(1 + 0.229573\;\ep)^{1.28098}} \; .
$$
In the fourth order $(k=4)$, we get
$$
f_4^*(\ep) = 2 (1 - 0.343467\; \ep)^{0.411916} (1 + 3.21435\; \ep)^{0.0051269} \;  .
$$
The fifth order $(k=5)$ yields
$$
f_5^*(\ep) =  2 (1 + \ep)^{0.0459476} (1 - 0.172432\;\ep)^{1.02395}  
(1 + 4.48424\; \ep)^{0.00125184} \;  .
$$
The sixth order $k=6$ gives
$$
f_6^*(\ep) = 2 (1 - 0.261332\; \ep)^{0.58267} (1 + 1.98816\; \ep)^{0.0126183} 
(1 + 5.44898\; \ep)^{0.000400602} \; .
$$  
And the approximant of seventh order $(k=7)$ is given by the expression
$$
f_7^*(\ep) =   2(1 + \ep)^{0.0275211} (1 - 0.222915\; \ep)^{0.734195}  
(1 + 3.13825\; \ep)^{0.0032879} (1 + 6.2879\; \ep)^{0.000131004} \; . 
$$
Substituting here $\ep=1$, we come to the corresponding values of the exponent $\nu$. 

All other factor approximants for the related expansions are obtained in the same way. 
In addition to the exponents $\nu$, $\eta$, and $\omega$, we find other exponents 
$\alpha$, $\beta$, $\gamma$, and $\delta$ through the relations
\be
\label{38}
 \al = 2 - 3\nu \; , \qquad \bt = \frac{\nu}{2}\; (1 +\eta) \; \, \qquad
\gm = \nu ( 2 -\eta) \; , \qquad \dlt = \frac{5-\eta}{1+\eta} \;  .
\ee
Table 7 summarizes these results.   

As is seen, the self-similar factor approximants are in good agreement with the 
results of other methods, while the approach using factor approximants is very 
simple and allowing for the explicit analytical construction of these approximants. 
  
It is worth mentioning one point, where the situation remains not completely settled.
This is the case of the exponent $\alpha$ for $N=2$. The case of $N=2$ is of special 
interest representing the superfluid helium and magnetic $XY$ models 
\cite{Kleinert_44,Shalaby_45}. The situation concerns the fact that practically all 
calculational methods yield the values of $\alpha$ that are in good accordance with 
each other as well as with the experimental values, but that are a bit lower than the 
result of Monte Carlo simulations and the value following from the conformal bootstrap 
conjecture. 

Table 8 illustrates the situation for $N=2$, comparing the exponent $\alpha$ obtained 
by different methods listed in Table 2 with experimental data, including the extremely 
precise results of the specific heat measurements for liquid helium in zero gravity 
\cite{Lipa_46} and the series of measurements for several magnetic materials of the 
$XY$ class \cite{Vasilev_47,Olega_48,Olega_49}.

\section{Discussion}
 
The method of self-similar transformations is a very simple and convenient tool 
for the summation of asymptotic series. The basis of the method is the consideration 
of the transfer from one approximation to another as a motion in the space of 
approximations, with the approximation order playing the role of time. The motion
in the vicinity of a fixed point is described by an equation having the form of a
self-similar relation, which is equivalent to a renormalization-group equation.    
The fixed point of the evolution equation defines the sought effective limit of the 
transformed sequence. A representation for the effective limit acquires the form
of self-similar factor approximants. Following the usual way of dealing with sequence
transformations, we show that the method is regular and then extrapolate it to the
case of divergent sequences.   

We apply the method of self-similar factor transformations for the summation of 
$\varepsilon$ expansions for the $O(N)$- symmetric theory in three dimensions. 
The series of seventh order in $\varepsilon$ are used. The method is shown to provide 
accurate approximations, at the same time being very simple and allowing for the 
construction of explicit analytical expressions. The results are compatible with
other known methods of summation.

Employing this method, it is even possible to get reasonable estimates for two-dimensional 
systems, for which one has to set $\varepsilon = 2$. For the two-dimensional systems, the 
symmetry can be broken only for $N = 0$ and $N = 1$ \cite{Pelissetto_50}. The estimates 
for the corresponding critical exponents can be compared with the values conjectured in 
Ref. \cite{Nienhuis_51} for $N=0$ and with the known exact values \cite{Calabrese_52}  
for $N=1$. Thus the self-similar factor approximants for $N=0$ give $\nu=0.748$, 
$\eta=0.18$, and $\om=1.62$, as compared with the exact values $\nu=0.75$, $\eta=5/24$, 
and $\om=2$. And for $N=1$, we obtain $\nu=0.997$, $\eta=0.21$, and $\om=1.6$, as compared 
with the exact values $\nu=1$, $\eta=0.25$, and $\om=1.75$. 

\vskip 5mm

{\bf Author Contributions}

Both the authors, V.I. Yukalov and E.P. Yukalova, equally contributed to this work.

\vskip 2mm

This research did not receive any specific grant from funding agencies in the public, 
commercial, or not-for-profit sectors.

\newpage

\begin{table}[ht]
\centering
\caption{Self-similar factor approximants of order $k=2,3,4,5,6,7$ for the exponent
$\nu$ and different number of components $N$.}
\vskip 3mm
\label{Table 1}
\renewcommand{\arraystretch}{1.25}
\begin{tabular}{|c|c|c|c|c|c|}  \hline 
$k$ &  $N=0$    &    $N=1$   &   $N=2$   &   $N=3$   &    $N=4$  \\ \hline
2   &  0.64688  &   0.73938  &  0.83405  &  0.91870  &   0.98471 \\ \hline
3   &  0.57949  &   0.61666  &  0.65176  &  0.68457  &   0.71492 \\ \hline
4   &  0.59026  &   0.63394  &  0.67562  &  0.71452  &   0.75006 \\ \hline
5   &  0.58665  &   0.62845  &  0.66875  &  0.70688  &   0.74221 \\ \hline
6   &  0.58789  &   0.63030  &  0.67127  &  0.71000  &   0.74573 \\ \hline
7   &  0.58744  &   0.62968  &  0.67073  &  0.70985  &   0.74614 \\ \hline
\end{tabular}
\end{table}

\begin{table}[ht]
\centering
\caption{Critical exponents for $N=0$, found by different methods: Self-similar 
factor approximants (FA), Monte Carlo simulations (MC), Conformal bootstrap (CB),
Hypergeometric Meijer summation (HGM), Borel summation with additional conjectures 
on the behaviour of coefficients (BAC), Borel summation with conformal mapping (BCM),
and Nonperturbative renormalization group (NPRG).}
\vskip 3mm
\label{Table 2}
\renewcommand{\arraystretch}{1.25}
\begin{tabular}{|c|c|c|c|}  \hline
$Method$ &  $\nu$         &     $\eta$     &    $\om$   \\  \hline
FA       &  0.5877 (2)    &   0.0301 (2)   &  0.821 (15)   \\ \hline
MC       &  0.5875970 (4) &   0.031043 (3) &  0.899 (12)  \\ \hline
CB       &  0.5877 (12)   &   0.0282 (4)   &    $-$      \\ \hline
HGM      &  0.5877 (2)    &   0.0312 (7)   &  0.8484 (17)   \\ \hline
BAC      &  0.5874 (2)    &   0.0304 (2)   &  0.846 (15)   \\ \hline
BCM      &  0.5874 (3)    &   0.0310 (7)   &  0.841 (13)   \\ \hline
NPRG     &  0.5876 (2)    &   0.0312 (9)   &  0.901 (24) \\ \hline
\end{tabular}
\end{table}

\begin{table}[ht]
\centering
\caption{Critical exponents for $N=1$, found by different methods listed in Table 2.}
\vskip 3mm
\label{Table 3}
\renewcommand{\arraystretch}{1.25}
\begin{tabular}{|c|c|c|c|}  \hline
$Method$ &  $\nu$         &     $\eta$     &    $\om$   \\  \hline
FA       &  0.6300 (3)    &   0.0353 (3)   &  0.808 (9)   \\ \hline
MC       &  0.63002 (10)  &   0.03627 (10) &  0.832 (6)  \\ \hline
CB       &  0.62999 (5)   &   0.03631 (3)  &  0.830 (2)     \\ \hline
HGM      &  0.6298 (2)    &   0.0365 (7)   &  0.8231 (5)   \\ \hline
BAC      &  0.6296 (3)    &   0.0355 (3)   &  0.827 (13)   \\ \hline
BCM      &  0.6292 (5)    &   0.0362 (6)   &  0.820 (7)   \\ \hline
NPRG     &  0.63012 (16)  &   0.0361 (11)  &  0.832 (14) \\ \hline
\end{tabular}
\end{table}

\begin{table}[ht]
\centering
\caption{Critical exponents for $N=2$, found by different methods listed in Table 2.}
\vskip 3mm
\label{Table 4}
\renewcommand{\arraystretch}{1.25}
\begin{tabular}{|c|c|c|c|}  \hline
$Method$ &  $\nu$         &     $\eta$     &    $\om$   \\  \hline
FA       &  0.6710 (3)    &   0.0372 (4)   &  0.809 (11)   \\ \hline
MC       &  0.67169 (7)   &   0.03810 (8)  &  0.789 (4)  \\ \hline
CB       &  0.67175 (10)  &   0.0385 (6)   &  0.811 (10)     \\ \hline
HGM      &  0.6708 (4)    &   0.0381 (6)   &  0.789 (13)   \\ \hline
BAC      &  0.6706 (2)    &   0.0374 (3)   &  0.808 (7)   \\ \hline
BCM      &  0.6690 (10)   &   0.0380 (6)   &  0.804 (3)   \\ \hline
NPRG     &  0.6716 (6)    &   0.0380 (13)  &  0.791 (8) \\ \hline
\end{tabular}
\end{table}

\begin{table}[ht]
\centering
\caption{Critical exponents for $N=3$, found by different methods listed in Table 2.}
\vskip 3mm
\label{Table 5}
\renewcommand{\arraystretch}{1.25}
\begin{tabular}{|c|c|c|c|}  \hline
$Method$ &  $\nu$         &     $\eta$     &    $\om$   \\  \hline
FA       &  0.7099 (1)    &   0.0372 (4)   &  0.7919 (3)   \\ \hline
MC       &  0.7116 (10)   &   0.0378 (3)   &  0.773      \\ \hline
CB       &  0.7121 (28)   &   0.0386 (12)  &  0.791 (22)     \\ \hline
HGM      &  0.7091 (2)    &   0.0381 (6)   &  0.764 (18)   \\ \hline
BAC      &  0.70944 (2)   &   0.0373 (3)   &  0.794 (4)   \\ \hline
BCM      &  0.7059 (20)   &   0.0378 (5)   &  0.795 (7)   \\ \hline
NPRG     &  0.7114 (9)    &   0.0376 (13)  &  0.796 (11) \\ \hline
\end{tabular}
\end{table}

\begin{table}[ht]
\centering
\caption{Critical exponents for $N=4$, found by different methods listed in Table 2.}
\vskip 3mm
\label{Table 6}
\renewcommand{\arraystretch}{1.25}
\begin{tabular}{|c|c|c|c|}  \hline
$Method$ &  $\nu$         &     $\eta$     &    $\om$   \\  \hline
FA       &  0.7459 (2)    &   0.0361 (4)   &  0.7913 (8)   \\ \hline
MC       &  0.750 (2)     &   0.0360 (3)   &  0.765 (30)   \\ \hline
CB       &  0.751 (3)     &   0.0378 (32)  &  0.817 (30)     \\ \hline
HGM      &  0.7443 (3)    &   0.0367 (4)   &  0.7519 (13)   \\ \hline
BAC      &  0.7449 (4)    &   0.0363 (2)   &  0.7863 (9)   \\ \hline
BCM      &  0.7397 (35)   &   0.0366 (4)   &  0.794 (9)   \\ \hline
NPRG     &  0.7478 (9)    &   0.0360 (12)  &  0.761 (12) \\ \hline
\end{tabular}
\end{table}

\begin{table}[ht]
\centering
\caption{Critical exponents for the three-dimensional $O(N)$-symmetric
$\vp^4$ field theory, calculated by means of self-similar factor approximants}
\vskip 3mm
\label{Table 7}
\renewcommand{\arraystretch}{1.25}
\begin{tabular}{|c|c|c|c|c|c|c|c|}  \hline
$N$ & $\al$     &  $\bt$    &  $\gm$    &  $\dlt$   &  $\nu$   &  $\eta$  &  $\om$   \\  \hline
0   & 0.2369    &  0.3027 &  1.15771  &  4.8247  &  0.5877  &  0.0301  &  0.821    \\ \hline
1   & 0.1100     &  0.3261 &  1.23776  &  4.7954  &  0.6300  &  0.0353  &  0.808  \\ \hline
2   & $-$0.0130  &  0.3480 &  1.31704  &  4.7848  &  0.6710  &  0.0372  &  0.809  \\ \hline
3   & $-$0.1297 &  0.3682 &  1.39339  &  4.7848  &  0.7099  &  0.0372  &  0.792  \\ \hline
4   & $-$0.2377 &  0.3864 &  1.46487  &  4.7910  &  0.7459  &  0.0361  &  0.791  \\ \hline
\end{tabular}
\end{table}

\begin{table}[ht]
\centering
\caption{Critical exponents $\al$ for $N=2$, found by different methods listed in 
Table 2, and experimental data.}
\vskip 3mm
\label{Table 8}
\renewcommand{\arraystretch}{1.25}
\begin{tabular}{|c|c|}  \hline
$Method$                      &  $\al$         \\  \hline
FA                            &  $-$0.0130 (8)     \\ \hline
MC                            &  $-$0.0151 (2)      \\ \hline
CB                            &  $-$0.0152 (3)         \\ \hline
HGM                           &  $-$0.0124 (12)      \\ \hline
BAC                           &  $-$0.0118 (6)      \\ \hline
BCM                           &  $-$0.0070 (30)      \\ \hline
NPRG                          &  $-$0.0148 (18)    \\ \hline
experiment \cite{Lipa_46}     &  $-$0.0127 (3)    \\ \hline
experiment \cite{Vasilev_47} &  $-$0.0130 (30)    \\ \hline
experiment \cite{Olega_48}    &  $-$0.0130 (10)    \\ \hline
experiment \cite{Olega_49}    &  $-$0.0130 (20)    \\ \hline
\end{tabular}
\end{table}

\end{document}